\documentstyle[pra,aps]{revtex}
\begin{document} 
\preprint{}
\thispagestyle{empty}
\draft
%\twocolumn[\hsize\textwidth\columnwidth\hsize\csname @twocolumnfalse\endcsname
\title{ Pointer states via Decoherence in a Quantum Measurement} 
\author{Anu Venugopalan\cite{email}}
\address {Physical Research Laboratory, 
Navrangpura,  \\ Ahmedabad - 380 009, INDIA} 
\maketitle
\begin{center}
(June 1999)
\end{center}

\begin{abstract}
We consider the interaction of a quantum system (spin-1/2) with a macroscopic
quantum apparatus (harmonic oscillator) which in turn is coupled to a 
bath of harmonic oscillators.  
Exact solutions of the Markovian Master equation
show that the reduced density matrix of the system-apparatus combine
decoheres to a statistical mixture where up and down spins eventually
correlate with pointer states of the  apparatus (harmonic oscillator),
with associated probabilities in accordance with quantum principles.
For the zero temperature bath these pointer states turn out to be 
{\em coherent states} of the harmonic oscillator (apparatus)
for {\em arbitrary initial states } of the apparatus. 
Further, we see that the
decoherence time is inversely proportional to the square of the separation
between the two coherent states with which the spins correlate.
For a high temperature bath pointer states no longer
remain coherent states but are Gaussian distributions  
({\em generalized coherent states}). Spin up and down states
of the system now correlate with 
{\em nearly diagonal distributions in position } of these
generalized coherent states.
The diagonalization in position increases with the
temperature of the bath. The off-diagonal elements in spin-space
decohere over a time scale which goes inversely as the square of the separation 
between the peaks of the two position distributions that correlate 
with the spin states.
Zurek's earlier approximate result for the
decoherence time is consistent with our exact results.
Our analysis brings out the importance of looking at a 
measurement-like-scenario where definite correlations are established
between the system and apparatus to determine the nature of the 
pointer basis of the apparatus. Further, our exact results 
demonstrate in an unambiguous 
way that the pointer states in this measurement model
emerge independent of the initial state of the apparatus.
\end{abstract}

%]

%\pagebreak
\pagestyle{plain}
\section{Introduction}
In a typical quantum measurement, the coupling between a microscopic
system and a macroscopic measuring apparatus results in an entangled 
state which seems to allow the read out of the apparatus (``meter") to
exist in a coherent superposition of macroscopically distinct
states, a situation which is difficult to reconcile with classical
intuition and perceptions. For a measurement to be 
classically interpretable and meaningful  
one expects the system-apparatus correlations to appear as a statistical
mixture. von Neumann\cite{von} postulated that an 
irreversible reduction process
takes such a quantum superposition (entanglement) 
into a statistical mixture in a 
measurement process. However, the apparent
nonunitary nature of such a  reduction raises several questions about the
validity of quantum mechanics and its connection with the emergence of
classicality. 

In recent years, decoherence\cite{zeh,zurek} is being widely discussed and
accepted as the mechanism responsible for the emergence of classicality
in a quantum measurement and the absence, in the real world, of 
Schr\"{o}dinger
Cat like states\cite{cat}. Decoherence results from the irreversible coupling
of  the apparatus to an environment. The appearance of classical behaviour
via decoherence in a quantum measurement-like scenario is marked by the 
dynamical transition of the reduced density matrix of the system-apparatus
combine from a pure entangled state to a statistical mixture with appropriate
correlations.  
This line of approach to the quantum measurement problem was initiated by
Zeh\cite{zeh} and later followed up by 
Zurek\cite{zurek} and several others.
Most studies relating to decoherence in the literature deal with an environment
modeled by a collection of harmonic oscillators with which the system of
interest interacts via a coordinate-coordinate coupling. The dynamics
of the reduced density matrix (after tracing over the degrees of freedom of 
the environment) is then described by the Markovian Master equation derived
separately 
by Caldeira and Leggett\cite{cl}, Agarwal\cite{gsa}, Dekker\cite{dek}
and others\cite{others} in the context of the quantum
Brownian motion. Using the Markovian Master equation with some
approximations, 
Zurek has argued that the density matrix for a free particle in 
an initial coherent superposition of two Gaussian
wavepackets separated by $\Delta x$ decoheres  (i.e., 
the off-diagonal elements of the density matrix decay) over a
 time scale which goes 
inversely as the
square of the separation ($\Delta x ^{2}$)  between the two parts of the
superposition\cite{zurek}. For classical systems and standard 
macroscopic separations, $\Delta x$, this `decoherence time' is shown to
be almost
$10^{-40}$ times smaller than the thermal relaxation time of the system.
Thus macroscopic superpositions are almost instantaneously reduced to
a statistical mixture\cite{zurek}. 
Savage and Walls\cite{sw} have solved the Master equation for
 a harmonic oscillator in an initial 
superposition of coherent states and seen the decay of the
off-diagonal elements of the density matrix at zero temperature and 
finite temperatures. The Master equation approach has been used by
Venugopalan et al\cite{av} to study a Stern-Gerlach type measurement model
where a spin-$1/2$ particle interacts with a quantum apparatus (represented
by the position 
and momentum degrees of freedom of the particle) which in turn is coupled
to a bath of oscillators through its position. They solve this equation
exactly and show that the reduced density matrix for the system and apparatus
is driven to a diagonal form as a consequence of
decoherence and the spin components correlate with momentum 
distributions\cite{av}. 

Recently, interest in the understanding of decoherence
has been heightened by advances on the experimental front. Brune et al
\cite{brune} experimentally created a mesoscopic superposition of
quantum states involving radiation fields with classically distinct phases
and observed its progressive decoherence to a statistical mixture through
two-atom correlation measurements.
Schrodinger cat-like states were recently created in an ion trap experiment
\cite {nist} using a single Berrylium ion and a combination of static
and oscillating electric fields. 
Though only a limited number of models have been approximately studied
so far, it is generally accepted now that the two main signatures 
of the decoherence mechanism are: (a) In the classical regime decoherence 
takes place over a time scale which is much smaller than the thermal
relaxation time of the system, and (b) the decoherence time goes inversely
as the square of the separation between the two parts of the superposition
\cite{zurek}. 
These features have been observed in the experiment of Brune et al\cite{brune}
and thus confirm the theoretical predictions\cite{zurek}.
Recently there have been several proposals to exploit 
purely quantum mechanical features like
the linear superposition principle and quantum 
entanglements\cite{az} to build high speed
quantum computers\cite{shor} and also to 
experimentally implement other ideas from 
quantum information
like quantum cryptography\cite{cryp} and quantum teleportation\cite{qt}. 
Since environmental influence is often unavoidable, decoherence can ruin
the functioning of such systems which rely heavily on the maintenance of 
quantum coherence. A clearer understanding of the behaviour of quantum
coherences in dissipative environments is, thus, of fundamental importance.
Our experience of the classical world suggests that unlike quantum systems,
which are allowed to exist in all possible states, classical systems only
exist in a few select states which are singled out by the environment
from a larger quantum menu\cite{zurek}.
These special states are the `preferred basis', also referred to as the
`pointer states' in a quantum-measurement-like scenario\cite{zurek}. In spite
of the progress in the theoretical and experimental understanding of 
decoherence, the models studied so far 
do not answer the question concerning the
nature of the preferred basis satisfactorily.
For simplified models where  the self Hamiltonian of the system has
either been ignored or considered co-diagonal with the interaction Hamiltonian, the `pointer' variable  has been shown to be the one which commutes 
with the interaction Hamiltonian\cite{zurek}. 
However, in more general situations where
all terms are included and the various parts of the Hamiltonian may not
commute, it is not obvious what decides the preferred basis. 
For the coordinate-coordinate coupling
model, the position basis is intuitively expected 
to emerge as the preferred basis.
However, this is contrary to the conclusion of Venugopalan et al\cite{av}
in their analysis of the Stern-Gerlach measurement model 
where the spin components eventually correlate with distributions which
are completely diagonal in the 
momentum basis and only approximately diagonal in the 
position basis\cite{av}.
In the literature, the preferred basis has been
variously described as the one in which the final state density matrix
becomes diagonal or that set of basis states which are characterized 
by maximum stability or a minimum increase in linear or statistical 
entropy, decided by a `predictability sieve'\cite{zhp}. In a
measurement-like scenario the pointer basis should be understood as
those states of the apparatus in which correlations with the
system states are eventually established {\em irrespective of the
initial states of the apparatus}. 
Using the Markovian Master equation for a harmonic oscillator coupled
to a heat bath and the criterion of the 
`predictability sieve' Zurek  argues that coherent 
states emerge as the preferred basis. In a recent paper, 
Paz and Zurek\cite{paz}
investigate decoherence in the limit of weak interaction with the
environment and show that the eigenstates of energy emerge as
pointer states. Roy and Venugopalan have recently 
obtained the exact solutions
of the Markovian Master Equation for a harmonic oscillator and a free
particle in a compact factorizable form and have shown that the density
matrix diagonalizes in the energy basis which is number states for the
oscillator and momentum states for the free particle for arbitrary
initial conditions\cite{smr}.  
It is intuitive that the pointer states should naturally be a consequence
of the interplay between the various components of the total Hamiltonian
and one should also expect them to be {\em  independent of the initial 
state of the system/apparatus}.
The limited number of examples studied so far 
do not bring out this
feature clearly.

In this paper we seek to analyze a measurement like scenario where
a spin-1/2 particle is coupled to a harmonic oscillator through its 
coordinate and the oscillator in turn is coupled to a bath of oscillators
via coordinate-coordinate coupling. 
The dynamics of the system-apparatus combine is studied via the Markovian
Master equation for zero temperature and high temperature cases. 
Exact solutions for arbitrary initial states of
the apparatus clearly show that the spin components eventually
correlate with {\em coherent states} of the apparatus at zero
temperature. This
brings out the role of coherent states as the pointer basis in an
unambiguous way.
At high temperatures the pointer states are Gaussian distributions
(generalized coherent
states) and up and down spins correlate with 
{\em nearly diagonal position distributions} of
these generalized coherent states. 
We also see the two main signatures of decoherence in
the measurement, i.e, the decoherence time is much shorter than the
thermal relaxation time in the classical regime
and it goes inversely as the square of the
separation between the `pointer states' with which the spin 
components correlate for zero temperature as well as for the high 
temperature case.  
The model we have considered is equivalent to the spin-boson Hamiltonian
and also corresponds to the physical example of a two-level atom 
coupled to a single
mode of a radiation field - a simple model that describes many
interesting physical situations in quantum optics\cite{jc} which could be
used to produce a system-apparatus entangled state where decoherence
can be experimentally monitored. 
For example, Brune et al\cite{brune} have used a 
Ramsey-type experiment using two-level Rydberg atoms 
and microwave cavities to
produce superposition states of the electromagnetic field
as well as atom-field entangled states which interact with a bath. 
Further, they have monitored the progressive 
decoherence of these pure states to a statistical mixture. 
Meekhof et al\cite{meek} have created thermal, Fock, coherent
and squeezed states of motion of a harmonically bound, cooled and trapped 
Beryllium ion 
where the coupling between 
its motion and internal states can be described
by the Jaynes Cummings-type interaction\cite{jc}. This, again, involves
a two-level atom radiatively coupled to the single mode radiation field. 
It is possible that similar systems, with suitable modifications
could be used to physically implement the 
system-apparatus entangled state of the model analyzed in this 
paper
and to subsequently monitor its decoherence mechanism.
The rest of the paper is organized as follows. In Section II we
introduce our model for the measurement and the equivalent Markovian
Master equation and solve it for the reduced density matrix of the
system-apparatus combine. In Section III we analyze our results
and observations and discuss the pointer basis.
Finally, 
in Section
IV we summarize the main results of this paper.

\section{Reduced density matrix for the system-apparatus } 

Consider our model for the measurement of spin. A spin-1/2 represents
the system. A harmonic oscillator represents the apparatus which is
coupled to the system via its coordinate. The harmonic oscillator
can be considered as a (macroscopic) measuring apparatus in 
the sense that it can measure
the state of the system (spin) via its position/momentum degrees of freedom,
which have well-understood classical distributions.
Alternately, this apparatus (harmonic oscillator)
could also be thought of as corresponding to
a single mode of a radiation field whose quantum
state correlates with the state of the system (spin) and hence can affect
a measurement of the spin. 
The apparatus is in turn 
coupled 
via a coordinate-coordinate coupling to a 
collection of oscillators which represent the environment.
The environmental influence via this bath of oscillators
brings about the decoherence of the entangled system-apparatus 
pure state to 
a statistical mixture.  
This arrangement represents {\em a general model for quantum 
measurement} and the total
Hamiltonian for such a system can be written as 
\begin{equation}
H= \frac{p^{2}}{2m} + \frac{1}{2}m \omega^{2} x^{2} + 
\lambda \sigma_{Z} + \epsilon x 
\sigma_{Z} + 
\sum_{j}  \frac{P_{j}^{2}}{2M_{j}} +
\frac{M_{j} \Omega_{j}^{2}}{2} \bigl ( X_{j}-\frac{C_{j}x}{M_{j} 
\Omega_{j}^{2}} \bigr )^{2}.
\end{equation}
Here $x$ and $p$ denote the position and momentum of the harmonic 
oscillator (apparatus) of mass 
$m$, and frequency $\omega$. $\lambda \sigma_{Z}$ is the  
the Hamiltonian of the system and $\epsilon$ is the 
strength of the system-apparatus coupling. The last term
represents the Hamiltonian for the bath of oscillators (environment) 
and the apparatus-environment
interaction.  
$X_{j}$ and $ P_{j}$ are the position and momentum coordinates of the
jth harmonic oscillator of the bath, $C_{j}$'s are the coupling strengths and
$\Omega_{j}$s are the frequencies of the oscillators comprising the
bath\cite{cg}.
For our analysis we deal directly with the
the Markovian Master equation for the reduced
density matrix for the system-apparatus combine in the
in the $|s,x \rangle$ representation where the environmental degrees of 
freedom have been traced out\cite{av}: 
\begin{eqnarray}
\frac{\partial \rho _{ss'}(x,x',t)}{\partial t} &=& \Big [ -\frac{\hbar}{2
i m} \bigl( \frac {\partial ^{2}}{\partial x^{2}} -
 \frac{\partial ^{2}}{\partial x'^{2}} \bigr) - 
\gamma (x-x ') \bigl( \frac {\partial}{\partial x} - \frac{\partial}
{\partial x'} \bigr) \\ \nonumber 
&& -  \frac{D}{4 \hbar^{2}}(x-x')^{2}  
- \frac{m \omega^{2}}{2 i \hbar}
(x^{2} - x'^{2}) 
+ \frac{i \epsilon(xs-x's')}{\hbar} + 
\frac{i \lambda (s-s')}{\hbar} 
\Big ]\rho_{ss'}
(x,x',t),
\end{eqnarray}
where $s,s' = +1 $ (or $\uparrow$ ) or $-1$ (or $\downarrow$).
Thus (2) represents four equations in the
coordinate representation, each corresponding to one of the
four elements
in spin space
($\uparrow \uparrow,\uparrow
\downarrow,\downarrow \uparrow,
\downarrow \downarrow $)
of the reduced density matrix for the system-apparatus combine\cite{av}.
Here $\gamma$ is the relaxation rate and 
\begin{equation}
D=8m\gamma \hbar \omega(\bar{n} + \frac{1}{2}),
\end{equation}
where
\begin{equation}
\bar{n}=\Big [ \exp{ (\hbar \omega / k_{B}T)} - 1 \Big ]^{-1},
\end{equation}
is the expected number of quanta in a harmonic oscillator of frequency
$\omega$ in equilibrium at temperature $T$ and 
$k_{B}$ is the Boltzmann coefficient\cite{gsa}. At $T=0$, $D=4m\gamma\omega
\hbar$ and for a high temperature bath, $D=8m\gamma k_{B}T$. 
Consider (2) in the changed coordinates:
\begin{equation}
R=\frac{x + x'}{2}, r= x-x'.
\end{equation}
The spin diagonal density matrix, $\rho_{d}$, and the spin off-diagonal
density matrix, $\rho_{od}$, obey the equations:
\begin{equation}
\frac{ \partial \rho_{d}(R,r,t)}{\partial t} =-\left [ \frac{\hbar}{im} 
\frac{\partial^{2}}{\partial r \partial R} + 2 \gamma
\frac{ \partial}{\partial r} + \frac{D r^{2}}{4 \hbar^{2}} +
\frac{m \omega^{2} r R}{i \hbar} \mp \frac{i \epsilon r}{\hbar}
\right ] \rho_{d}(R,r,t),
\end{equation}
where the $`+'$ sign in the last term corresponds to $\rho_{\uparrow 
\uparrow}$ and $`-'$ to $\rho_{\downarrow \downarrow}$, and
\begin{equation}
\frac{\partial \rho_{od}(R,r,t)}{\partial t}=-\left [ \frac{\hbar}{im}
\frac{\partial^{2}}{\partial r \partial R} + 2 \gamma
\frac{ \partial}{\partial r} + \frac{Dr^{2}}{4 \hbar^{2}} +
\frac{m\omega^{2}rR}{i\hbar} \mp \frac{2 i\epsilon R}{\hbar}
\mp \frac{2i\lambda}{\hbar} \right ] \rho_{od}(R,r,t),
\end{equation}
where the upper signs in the last two terms correspond to $\rho_{\uparrow
\downarrow}$ and the lower ones to $\rho_{\downarrow \uparrow}$. To solve
these equation it is convenient to take a partial Fourier transform in the
variable $R$:
\begin{equation}
\rho(Q,r,t)= \int_{-\infty}^{\infty} 
e^{iQR} \rho(R,r,t)dR.
\end{equation} 
Equations (6) and (7) then simplify to a pair of first-order partial 
differential equations:
\begin{equation}
\frac{\partial \rho_{d}(Q,r,t)}{\partial t}= \left ( \frac{\hbar Q}{m} 
- 2 \gamma r \right ) \frac{\partial 
\rho_{d}}{\partial r} - \frac{m \omega^{2} r}{\hbar} \frac{\partial \rho_{d}}
{\partial Q} - \frac{Dr^{2}}{4 \hbar^{2}} \rho_{d}
\pm \frac{i \epsilon r}{\hbar} \rho_{d},
\end{equation}
\begin{equation}
\frac{\partial \rho_{od}(Q,r,t)}{\partial t}= \left ( \frac{\hbar Q}{m} 
- 2 \gamma r \right ) 
\frac{\partial
\rho_{od}}{\partial r} - \frac{m \omega^{2} r}{\hbar} \frac{\partial \rho_{od}}
{\partial Q} - \frac{Dr^{2}}{4 \hbar^{2}} \rho_{od}
\pm \frac{2 \epsilon}{\hbar} \frac{\partial \rho_{od}}{\partial Q}
\pm \frac{2i\lambda}{\hbar} \rho_{od}.
\end{equation}
Equations (9) and (10) can be solved by the method of characteristics
\cite{av,smr,ch}.
Let the initial state of the system-apparatus combination 
be a product
of any arbitrary state of the apparatus (oscillator) 
and a general superposition
state for the spin-1/2 system of the form:

\begin{equation}
\psi(x,s,0)=\phi(x) \otimes \Big [ a|\uparrow \rangle + b|\downarrow \rangle
\Big ],
\end{equation}
where $\phi(x)$ is {\em any} initial state of the harmonic oscillator 
(apparatus). The time evolved density matrix would then appear as:
\begin{eqnarray}
\rho & = & |a|^{2} |\uparrow \rangle \langle \uparrow| 
\hspace*{0.2cm} \rho_{\uparrow \uparrow} (x, x',t) +
|b|^{2} |\downarrow \rangle \langle \downarrow | 
 \hspace*{0.2cm} \rho_{ \downarrow \downarrow}
(x, x',t) \\ \nonumber
&& + ab^{*} |\uparrow \rangle \langle \downarrow| 
\hspace*{0.2cm}\rho_{ \uparrow \downarrow} (x, x',t) 
+ a^{*}b |\downarrow \rangle 
\langle \uparrow | \hspace*{0.2cm} \rho_{ \downarrow \uparrow} (x, x',t).
\end{eqnarray}
The solution for the spin off-diagonal elements of the density matrix 
(corresponding to $\rho_{ \uparrow \downarrow}$ and 
$\rho_{ \uparrow \downarrow}$ ) 
in the partial Fourier transform representation is:
\begin{equation}
\rho_{od}(Q,r,t)= \rho_{od}(Q',r',0) \exp \left \{ - \frac{ \epsilon^{2} t D}
{\hbar^{2} m^{2}\omega^{4} }  \mp \frac{\epsilon D}{m \omega^{2}\hbar^{2}}Z_{1}
- \frac{D}{4\hbar^{2}}Z_{2} \pm \frac{2 i \lambda t}{\hbar} \right \},
\end{equation}
where

\begin{eqnarray}
Z_{1}&=& 
\frac{m \lambda_{+} \Gamma}{\hbar} \left (
 Q -\frac{r}{\lambda_{+}} \pm \frac{2 \epsilon}{\hbar \omega^{2}} ( 2\gamma
- \frac{\hbar}{m \lambda_{+}}) \right )  (1-e^{-\frac{\hbar t}{m \lambda_{+}}}
) \\ \nonumber
&& + \frac{m \lambda_{-} \Gamma }{\hbar} \left (
Q-\frac{r}{\lambda_{-}} \mp \frac{2 \epsilon}{\hbar \omega^{2}} ( 2 \gamma
- \frac{\hbar} {m \lambda_{-}} ) \right )  (1 - e^{-\frac{\hbar t}
{m \lambda_{-}}} ) , \\ \nonumber 
Z_{2}&=&  
\frac{m \Gamma^{2}  \lambda_{+}}{2 \hbar} \left (
 Q -\frac{r}{\lambda_{+}} \pm \frac{2 \epsilon}{\hbar \omega^{2}} ( 2\gamma
- \frac{\hbar}{m \lambda_{+}}) \right )^{2} ( 1 - e^{- \frac{2 \hbar t}{m
\lambda_{+}}})  \\ \nonumber
&&  + \frac{m  \Gamma^{2}  \lambda_{-}}{2 \hbar} \left ( 
Q-\frac{r}{\lambda_{-}} \mp \frac{2 \epsilon}{\hbar \omega^{2}} ( 2 \gamma
- \frac{\hbar} {m \lambda_{-}} ) \right )^{2} ( 1 - e^{-\frac{2 \hbar t}{m
\lambda_{-}}}) \\ \nonumber 
&& - \frac{ \Gamma^{2} }{\gamma} \left (
 Q -\frac{r}{\lambda_{+}} \pm \frac{2 \epsilon}{\hbar \omega^{2}} ( 2\gamma
- \frac{\hbar}{m \lambda_{+}}) \right ) 
\left (
Q-\frac{r}{\lambda_{-}} \mp \frac{2 \epsilon}{\hbar \omega^{2}} ( 2 \gamma
- \frac{\hbar} {m \lambda_{-}} ) \right ) (1- e^{-2\gamma t}), \\ \nonumber
\end{eqnarray}
and
\begin{eqnarray}
\Gamma&=& \frac{\lambda_{+} \lambda_{-}}{\lambda_{+} - \lambda_{-}}, \\ 
\nonumber
\lambda_{\pm}&=&\frac{\hbar}{m\omega^{2}}
(\gamma \pm \sqrt{\gamma^{2}-
\omega^{2} } ). \\ \nonumber 
\end{eqnarray}
$Q'$ and $r'$ are defined as:
\begin{eqnarray}
Q'&=& \frac{ c_{+}\lambda_{+} - c_{-}\lambda_{-}}{\lambda_{+} -
\lambda_{-}} \pm \frac{4 \epsilon \gamma}{\hbar \omega^{2}},
\\ \nonumber
r'&=& \Gamma (c_{+} - c_{-}) \pm \frac{2 \epsilon}{m \omega^{2}},
\end{eqnarray}
where the coefficients $c_{\pm}$ are:
\begin{eqnarray}
c_{+}&=& \left (
 Q -\frac{r}{\lambda_{+}} \mp \frac{2 \epsilon}{\hbar \omega^{2}} (
2\gamma
- \frac{\hbar}{m \lambda_{+}})\right ) e^{-\frac{\hbar t}{m
\lambda_{+}}} , \\ \nonumber 
c_{-}& =& \left (
Q-\frac{r}{\lambda_{-}} \mp \frac{2 \epsilon}{\hbar \omega^{2}} ( 2
\gamma
- \frac{\hbar} {m \lambda_{-}} )\right ) e^{-\frac{\hbar
t}{m \lambda_{-}}}. 
\end{eqnarray}
It is clear from (13) that the leading order decay term for the
spin off-diagonal elements of the reduced density matrix goes as
$e^{-\alpha t}$ which
would drive the entire expression to zero with time, independent of
all other arguments. This, essentially, is the decoherence of the pure
state density matrix and happens over a time scale:
\begin{equation}
\tau_{D}= \frac{ \hbar^{2} m^{2} \omega^{4}}{D \epsilon^{2}}.
\end{equation}
We will discuss the features of this decoherence in greater detail in
the next section. Consider now the solution for the spin diagonal 
elements of the reduced density matrix for the system-apparatus:
\begin{equation}
\rho_{d}(Q,r.t)=\rho(Q'',r'',0)\exp{ \left ( - \frac{D}{4 \hbar^{2}}Z_{3} 
\pm \frac{i \epsilon}{\hbar}Z_{4} \right )},
\end{equation}
where
\begin{eqnarray}
Z_{3} &=&
\frac{m \Gamma^{2}  \lambda_{+}}{2 \hbar} (
 Q -\frac{r}{\lambda_{+}})^{2} 
( 1 - e^{- \frac{2 \hbar
t}{m
\lambda_{+}}})  \\ \nonumber
&&  + \frac{m  \Gamma^{2}  \lambda_{-}}{2 \hbar}  (
Q-\frac{r}{\lambda_{-}})^{2} 
( 1 - e^{-\frac{2 \hbar t}{m \lambda_{-}}}) \\ \nonumber 
&& - \frac{ \Gamma^{2} }{\gamma} (
 Q -\frac{r}{\lambda_{+}})
(Q-\frac{r}{\lambda_{-}})
(1- e^{-2\gamma t}), \\ \nonumber
Z_{4}&=& \frac{m \Gamma \lambda_{+}}{\hbar} 
( Q -\frac{r}{\lambda_{+}})(1-e^{\frac{\hbar t}{m\lambda_{+}}}) - \\ \nonumber
&& \frac{m \Gamma \lambda_{-}}{\hbar} (Q -\frac{r}{\lambda_{-}})
(1-e^{\frac{\hbar t}{m\lambda_{-}}}),   
\end{eqnarray}
and $Q''$ and $r''$ are defined as:
\begin{eqnarray}
Q''&=& \frac{ c'_{+}\lambda_{+} - c'_{-}\lambda_{-}}{\lambda_{+} -
\lambda_{-}},
\\ \nonumber
r''&=& \Gamma (c'_{+} - c'_{-}).
\end{eqnarray}
The coefficients $c'_{\pm}$ are:
\begin{eqnarray}
c'_{\pm}& =& \left (
Q-\frac{r}{\lambda_{-}} 
\right ) e^{-\frac{\hbar
t}{m \lambda_{\pm}}}.
\end{eqnarray}
(13) and (19)  are the exact solutions 
corresponding to the two diagonal and two off-diagonal elements in
spin space of the reduced density matrix of the system-apparatus in
the $Q,r$ representation.  
In the next section we analyze these solutions and discuss the
decoherence mechanism and the emergence of the  pointer basis in this
measurement model. 

\section{Decoherence and Preferred Basis}
In the previous section we have seen (from Eq. (13)) that the spin 
off-diagonal elements
of the reduced density matrix of the system-apparatus {\em decay to
zero} with time irrespective of the initial state of the apparatus. 
Thus the entangled system-apparatus 
pure state (12) eventually diagonalizes over a time scale given
by (18) to a {\em mixed density matrix} with definite system-apparatus
correlations:
\begin{equation}
\rho =  |a|^{2} |\uparrow \rangle \langle \uparrow|
\hspace*{0.2cm} \rho_{\uparrow \uparrow} (x, x',t) 
+ |b|^{2} |\downarrow \rangle \langle \downarrow |
 \hspace*{0.2cm} \rho_{ \downarrow \downarrow}
(x, x',t).
\end{equation}
The environmental influence manifested via the Markovian Master
equation  for the reduced density matrix has, thus, 
clearly destroyed the off-diagonal elements in spin-space and
affected a measurement of the spin.
Let us now examine the nature of the pointer states, 
$\rho_{\uparrow \uparrow}$ and $\rho_{ \downarrow \downarrow}$, which
correlate with up and down spin states in (23). 
In the $Q,r$ representation
this is given by Eq. (19). Consider the long time limit
($t\rightarrow \infty$) of (19) ($ \gamma > \omega$). It can be seen
that at long times (19) takes the form:
\begin{eqnarray}
\rho_{d}(Q,r,t)&=&\rho(0,0,0) \exp \left \{ -\frac{D}{16
m^{2} \omega^{2} \gamma} \left ( Q^{2} + \frac{m^{2} \omega^{2}
r^{2}}{\hbar^{2}} \right ) \pm \frac{i\epsilon Q}{m\omega^{2}} \right
\} \\ \nonumber
&=& \frac{1}{2\pi} \exp \left \{ -\frac{D}{16
m^{2} \omega^{2} \gamma} \left ( Q^{2} + \frac{m^{2} \omega^{2}
r^{2}}{\hbar^{2}} \right ) \pm \frac{i\epsilon Q}{m\omega^{2}} \right
\}. 
\end{eqnarray} 
The Fourier transform of (24) in the position
representation  $(x, x')$ is:
\begin{eqnarray}
\rho_{d}(R, r, t)&=& 2 m \omega \sqrt{\frac{\gamma}{\pi D}}
\exp{ \left \{- \frac{4 m^{2} \omega^{2} \gamma}{D} \left (R \pm
\frac{\epsilon}{m\omega^{2}} \right )^{2}
 - \frac{D r^{2}}{16 \hbar^{2} \gamma} 
\right\}} \\ \nonumber
&=& 2 m \omega \sqrt{\frac{\gamma}{\pi D}} \exp{ \left \{- \frac{4
m^{2} \omega^{2} \gamma}{D} \left ( \frac{x+x'}{2} \pm 
\frac{\epsilon}{m\omega^{2}}
\right )^{2} - \frac{D (x-x')^{2}}{16 \hbar^{2} \gamma}
\right\}}.
\end{eqnarray} 
(25) is the final form of the pointer states which eventually correlate
with up and down spins in the mixed density matrix (23).
\subsection{Zero Temperature}
For the zero temperature Markovian bath, $D=4m\omega \gamma \hbar$.
Substituting for $D$ in (25) gives:
\begin{equation}
\rho_{d}(R,r,t)=\sqrt{\frac{m \omega}{\pi \hbar}} \exp{ \left \{-
\frac{m\omega}{\hbar} \left (R \pm
\frac{\epsilon}{m\omega^{2}} \right )^{2}
 - \frac{m \omega r^{2}}{4 \hbar} \right\}}.
\end{equation}
This is nothing but the density matrix corresponding to a {\em coherent 
state}, $|\alpha \rangle$, 
of a harmonic oscillator with zero
mean momentum, mean positions $ = \pm \epsilon/m\omega^{2}$,
and
\begin{equation}
|\alpha|^{2}=\frac{m \omega}{2 \hbar} \left
(\frac{\epsilon}{m\omega^{2}} \right ) ^{2} = \frac{\epsilon ^{2}}{2 m 
\omega ^{3} \hbar}.
\end{equation}
Thus up and down spins correlate with {\em coherent states} which clearly
establishes the coherent states as the pointer basis or the preferred
states here. 
Zurek et al\cite{zhp} have earlier derived an approximate 
expression for the `predictability sieve' which is the measure of the
increase in entropy, $S=Tr(\rho-\rho^{2})$, for a harmonic oscillator
coupled to a heat bath whose dynamics is described by the Markovian
Master equation. If  $\Delta x$ and $\Delta p$ are the initial
dispersions in $x$ and $p$ , in the limit of weak coupling 
and under the assumption that the
initial state remains approximately pure, they show that 
\begin{equation}
\frac{dS}{dt} \sim 4D\Delta x^{2}.
\end{equation}
Further, they integrate (28) in the weak coupling limit 
over an oscillator period, after
replacing the free Heisenberg equations for the oscillator operators
and show that
\begin{equation}
S(t)=2D\left[\Delta x^{2} + \frac{\Delta p^{2}}{m^{2} \omega^{2}}
\right].
\end{equation}
The
quantity (29) is minimum if $\Delta x \Delta p=\hbar/2 $, and $\Delta
x^{2} = \hbar/2 m \omega$. This corresponds to the spread in position
of the ground state or of  a coherent state of an oscillator. On this
basis Zurek et al claim that the coherent states are the preferred
basis for a harmonic oscillator\cite{zhp}. 
Our analysis of the exact solutions
for the full Master equation here shows  
in a more rigorous way that coherent states emerge naturally as
the apparatus states that eventually correlate
with the system states. Moreover, this happens for 
 {\em arbitrary initial states of the apparatus}
which firmly establishes the fact that coherent 
states are truly the preferred 
states for the apparatus in this measurement model. 
One can see
that the decoherence time (18) is:
\begin{equation}
\tau_{D}= \tau_{R} \left ( \frac{m
\omega ^{3} \hbar}{\epsilon ^{2}}
\right )
 = \frac{\tau_{R}}{ 2|\alpha|^{2}},
\end{equation}
where $\tau_{R} = \gamma^{-1}$ is the thermal relaxation time.  
When $|\alpha|^{2} \gg 1$, it is clear that $\tau_{D} \ll \tau_{R}$.
From (27) one can see that this would be the case
when one has conditions expected in the classical limit, i.e.,
when Planck's constant, $\hbar$, is small relative
to the actions involved.
It is in such a regime
that one would expect a fast decoherence of the  
superposition of `macroscopically distinct' pointer positions
to a statistical mixture. 
The main signatures of a quantum measurement
via the decoherence mechanism thus are clearly seen
here, namely: (a) In the classical regime there 
is a fast decoherence of the off-diagonal
elements of the spin density matrix over a time scale $\tau_{D}$ 
given by (30) which is {\em much smaller than} $\tau_{R}$, the
relaxation
time of the bath, 
(b) there is a {\em one-to-one correlation between the
spin states and the pointer states of the apparatus} which are {\em coherent
states}, and  (c) $\tau_{D}$ is inversely proportional to the
square of the separation $|\alpha|^{2}$ between the two
pointer states. The strength of this model is that the
system-apparatus correlations established at long times are 
permanent and the emergent pointer basis is  
independent of the initial state of the
apparatus.  

\subsection{High Temperature}
For a the high temperature thermal bath, $D=8 m \gamma k_{B}T$.
Substituting this in (25) gives:
\begin{equation}
\rho_{d}(R,r,t)=\sqrt{\frac{m \omega^{2}}{2 \pi k_{B}T}}
\exp{ \left \{ -\frac{m \omega^{2}}{2 k_{B}T} \left (
R \pm
\frac{\epsilon}{m\omega^{2}} \right )^{2}
-\frac{m k_{B}T r^{2}}{2 \hbar^{2}} \right\}}.
\end{equation}
(31) is no longer a coherent state but a Gaussian distribution which  
is also referred to as a  {\em generalized coherent state}. Tegmark
and Shapiro \cite{ts} have earlier shown that generalized coherent
states tend to be produced naturally when one looks at the reduced
Wigner distribution of infinite systems of coupled harmonic oscillators
at $t \rightarrow \infty$. Our results are in tune with their
predictions.
For high temperatures, on can see that the variance corresponding to
the off-diagonals elements in the position basis, $r$, is  small and
decreases with increasing temperature. Thus for a high temperature
bath, this generalized coherent state (31) is nearly diagonal in the
position representation. Spin up and down states are clearly then
correlated with these approximately diagonal position distributions
which are centered around $\pm \epsilon / m\omega^{2}$: 
\begin{eqnarray}
\rho & \sim &  |a|^{2} |\uparrow \rangle \langle \uparrow|
\hspace*{0.2cm} \otimes \sqrt{\frac{m \omega^{2}}{2 \pi k_{B}T}}
\exp{ \left \{ -\frac{m \omega^{2}}{2 k_{B}T} \left (
x + 
\frac{\epsilon}{m\omega^{2}} \right )^{2} \right \}} \\ \nonumber
&& + |b|^{2} |\downarrow \rangle \langle \downarrow |
 \hspace*{0.2cm} \otimes \sqrt{\frac{m \omega^{2}}{2 \pi k_{B}T}}
\exp{ \left \{ -\frac{m \omega^{2}}{2 k_{B}T} \left (x -
\frac{\epsilon}{m\omega^{2}} \right )^{2} \right \}}. 
\end{eqnarray}
It can be checked that spin-apparatus correlations do not exist in the
diagonal elements of the momentum basis and hence {\em position}
 is quite obviously
`preferred' by the environment. This contrasts with the Stern-Gerlach
model analyzed by Venugopalan et al\cite{av} where it is the momentum
distributions with which the spin-apparatus correlations ultimately
get established. The decoherence time (18) over which the spin off-diagonal
elements of the reduced density matrix of the system-apparatus decay for 
the high temperature
bath is now given by:
\begin{equation}
\tau_{D}' = \tau_{R} \left ( \frac{m \hbar^{2} \omega^{4}}{8
\epsilon^{2} k_{B}T} \right) 
= \tau_{R} \left (\frac{\lambda_{d}}{\Delta^{2}} \right
)^{2},
\end{equation}
where  $\lambda_{d}= \hbar/\sqrt{2mk_{B}T}$ is the thermal de Broglie
wavelength of the particle and $ \Delta= 2 \epsilon/ m\omega^{2}$ is
the separation between the peaks of the two pointer distributions in
(32). It is obvious that whenever $\Delta \gg \lambda_{d}$, there
is a fast decoherence of the entangled system-apparatus pure state to
a statistical mixture. Such a condition would correspond to a regime 
expected in 
the classical limit. 
Thus, once again, we can
clearly see the main signatures of the
decoherence mechanism here as in the zero temperature case discussed
above, namely, the off-diagonal elements in spin-space decohere to zero
completely in a time scale which is much smaller that $\gamma^{-1}$ 
and which goes inversely as the square of the
spatial separation between the `pointers', eventually leading to a
mixed density matrix (32) with appropriate system-apparatus correlations.  
The decoherence time (33) was earlier 
obtained by Zurek\cite{zurek} from the high temperature Markovian Master
equation for a free particle in an initial coherent superposition of
two Gaussian wavepackets separated by $\Delta$ 
under the approximation that the only
dominant term is $ \frac{D}{4 \hbar^{2}}(x-x')^{2}$, in (2). 
We obtain the same result for the decoherence time 
from {\em exact solutions of the full Master
equation} where no terms are neglected and no approximations 
are made. Moreover, this behaviour of
$\tau_{D}$ is
seen for all times unlike previous estimates of decoherence times
where decoherence was looked for at short times (in the limit of 
negligible friction).
Our results are also consistent with the solutions obtained by
Savage and Walls\cite{sw}
for a harmonic oscillator in an initial superposition of 
coherent states for the zero temperature and high temperature cases.
For the measurement model analyzed
in this paper, our exact solutions show that 
the final mixed state density matrix carrying
system-apparatus correlations is independent of the initial state of
the apparatus.
From (30) and (33) it is clear that the decoherence time for a high
temperature heat bath is much shorter than that for the zero temperature
bath ($\tau_{D}'/\tau_{D} \sim \hbar \omega /K_{B}T$)  and decreases
with
increase in temperature of the bath. Similar features have been seen 
by Kim and Buzek in their study of the influence of a heat bath on
superposition states of  light in a microwave cavity\cite{kb}. 

It is interesting to note that
the decoherence times (30) and (33) are
directly proportional to the mass and frequency of the apparatus 
(oscillator) and are longer for heavier oscillators with higher
frequencies. Of course, correspondingly, the `separation' between the
two `pointers' positions, $2 \epsilon / m \omega^{2}$, will be
smaller and hence their superpositions would decohere slower. Thus,
the `bigger' the cat-state, the faster the decoherence.  
In the experiment of Brune et al\cite{brune}, an
entangled state of the atom plus field (`meter') is formed and its
progressive decoherence to a statistical mixtures is 
monitored. However, since 
the fields eventually relax towards
vacuum, in their system 
the one-to-one correlation between the atom and meter
states are eventually lost. The  experimental study of
decoherence is thus confined to extremely short-time scales
which are much smaller than the cavity relaxation times\cite{brune}.
It is interesting to
note that in a physical realization of the measurement 
model discussed in this
paper, the system-apparatus correlations will persist for all times
for both the zero temperature and high temperature cases as is evident
from (26) and (32). The `permanence' of these system-apparatus
correlations, thus, makes this model very interesting to investigate
experimentally from the point of view of quantum measurement and
decoherence.

\section{Conclusions}
In this paper, we have investigated a quantum 
measurement model comprising of a
spin-1/2 (system), a harmonic oscillator (apparatus) and a bath of
oscillators (environment). Our interest has been to look at the exact
solutions for the 
dynamics of the reduced density matrix of the system and apparatus
via the Markovian Master equation which we have studied 
for the zero temperature
and high temperature cases.    
We show that the coupling of the apparatus to the environment leads to
the decoherence of the pure system-apparatus entangled state to a
statistical mixture with definite system-apparatus correlations, thus
affecting a measurement of the spin state. For both the zero temperature
and high temperature cases our exact solutions clearly demonstrate the
two main signatures of the decoherence mechanism in a quantum measurement, 
namely, (a) the
decoherence time is much smaller than the thermal relaxation time, 
and (b) the decoherence time is inversely proportional to
the square of the `separation' between the two  `pointers' that
correlate with the system states. Decoherence is much faster in a high
temperature bath compared to the zero temperature bath.
Our exact solutions also clearly show that
the final apparatus states with which the system states eventually correlate 
at long times (the `pointer states') 
are coherent states for the zero temperature bath and
nearly diagonal position distributions 
of a generalized coherent state
for the high
temperature heat bath.  The strength of this model and analysis is that 
it clearly demonstrates that the emergent pointer basis in
a measurement process is 
independent of the initial states of the apparatus.
This fact was intuitively obvious, but has not been shown so far in
the limited number of models studied in the literature. 
Our analysis also highlights the need to consider a measurement like
scenario to address the issue of the emergent pointer basis. The model
considered here is fairly generic and our exact solutions make this model
an interesting candidate to explore experimentally in the context of 
decoherence and 
quantum measurements.

\end{document}